\documentclass{JAC2003}

\usepackage{graphicx}
\usepackage{booktabs}
\usepackage{multicol}
\usepackage{amsmath}
\usepackage{url}

\begin{document}

\title{Two-dimensional models of the magnetic-field enhancement at pit and bump}

\author{Takayuki Kubo\thanks{kubotaka@post.kek.jp}, \\
KEK, High Energy Accelerator Research Organization, Tsukuba, Ibaraki 305-0801, Japan}

\maketitle

\begin{abstract}
Formulae of magnetic field enhancement at a two-dimensional semi-elliptical bump and a two-dimensional pit with chamfered edges are derived by using the method of conformal mapping. 
The latter can be regarded as an approximated model of the two-dimensional pit with round edges studied in Ref.~\cite{Kubo}. 
\end{abstract}

\section{Introduction}\label{section:introduction}

A surface magnetic-field of superconducting accelerating cavity could be enhanced at a defect on a cavity surface. 
The enhanced magnetic-field can be written as
\begin{eqnarray}
H({\bf r})=\beta({\bf r}) H_0 \hspace{0.8cm} (0\le \beta({\bf r}) \le \beta^*) \,, 
\end{eqnarray}
where 
$H_0$ is a surface magnetic-field of the cavity, 
$\beta({\bf r})$ is a magnetic-field enhancement factor at a position ${\bf r}$,  
and $\beta^*$ is the maximum value of $\beta({\bf r})$ along the defect.  
When $\beta H_0$ exceeds a threshold value, vortices start to penetrate into a superconductor 
and a surface dissipation increases, 
which could be a cause of quenches and limit the maximum accelerating field.
Understanding a relation between $\beta$ and the geometry of defect is a first step to understand the mechanism of quench at defect.

In Ref.~\cite{Kubo}, the two-dimensional (2D) pit with a triangular section is proposed as the minimum model of the pit, which contains only two parameters: 
a ratio of a curvature radius to a half-width of pit and a slope angle. 
First the analytical formula of $\beta(r)$ of a model with sharp edges was derived by the method of conformal mapping, 
from which that of a model with round edges was obtained by a polynomial extrapolation. 
The model was consistent with experiments~\cite{Kubo}. 
Note that the 2D well is a special case of this model. 
The functional form of the $\beta$ of the 2D well is also derived analytically in Ref.~\cite{Kubo}, 
which is consistent with the numerical result of 3D well obtained by Shemelin and Padamsee~\cite{shemelin}.

In this paper the method of conformal mapping is applied to other interesting geometry: 
a 2D semi-elliptical bump and a 2D pit with chamfered edges. 
The former is a good exercise of this method before we tackle the complicated problem. 
The latter can be regarded as an approximation of the 2D pit with round edges given in Ref.~\cite{Kubo}.

\section{Formulation}

The solution of a two-dimensional magnetostatics problem $(H_x, H_y)$ is given by 
\begin{eqnarray}
H_x - i H_y  = -\Phi'(z)  ,  \label{eq:Hx-iHy1}
\end{eqnarray}
where $\Phi(z)$ is an holomorphic function of a complex variable $z=x+iy$. 
When a complex potential $\widetilde{\Phi}(w)$ on a complex $w$-plane is known,  
$\Phi(z)$ is derived from $\widetilde{\Phi}(w)$ through a conformal mapping, $z=F(w)$, 
and is given by 
\begin{eqnarray}
\Phi(z) = \widetilde{\Phi}(F^{-1}(z)) \,, \label{eq:Phiz}   
\end{eqnarray}
where $F^{-1}$ is an inverse function of $F$. 
Then the $\beta$ can be computed as
\begin{eqnarray}
\beta(x,y) 
\!=\! \frac{|H_x \!-\! i H_y|}{H_0} = \frac{|\Phi'(z)|}{H_0} 
\!=\! \frac{1}{H_0} \biggl| \frac{\widetilde{\Phi}'(w)}{dF/dw} \biggr|   
,  \label{eq:beta}
\end{eqnarray}
where $dF^{-1}/dz = dw/dz= (dz/dw)^{-1} = (dF/dw)^{-1}$ is used. 
Eq.~(\ref{eq:beta}) is the general formula to obtain $\beta$ used below. 
Our tasks are to find $\widetilde{\Phi}(w)$ and $F(w)$ of each problem.

\section{2D semi-elliptic bump}\label{Formulation}

%
\begin{figure}[tb]
   \begin{center}
   \includegraphics[width=0.8\linewidth]{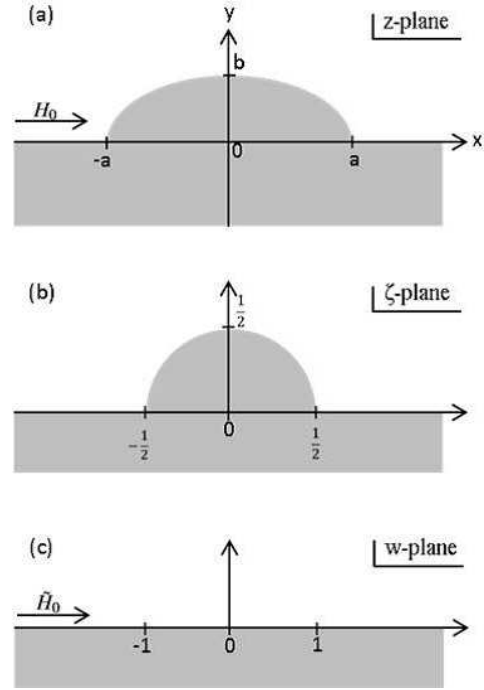}
   \end{center}\vspace{-0.2cm}
   \caption{
(a) 2D semi-elliptic bump with a major axis, $a$, and a minor axis, $b$, and its maps on (b) the $\zeta$-plane and (c) the $w$-plane. 
Gray regions correspond to superconductors in the Meissner state.  
A surface magnetic field far from the bump is given by $(H_x,H_y)=(H_0,\, 0)$ and $(H_u,H_v)=(\widetilde{H}_0,0)$ on the $z$- and $w$-plane, respectively. 
   }
\label{fig1}
\end{figure}

Fig.~\ref{fig1}(a) shows a semi-elliptical bump with a major axis, $a$, and a minor axis, $b$. 
Fig.~\ref{fig1}(b) and (c) are schematic views of its maps on $w$- and $\zeta$-plane, respectively.  
The map between the $w$- and $\zeta$-plane is given by
\begin{eqnarray}
\zeta = \frac{w+ \sqrt{w^2-1}}{2} \,.  \label{eq:bump_zeta-w}
\end{eqnarray}
and that between the $\zeta$- and $z$-plane is given by
\begin{eqnarray}
z = \frac{r_0}{2} \biggl( \frac{2\zeta}{e^{-\theta}} + \frac{e^{-\theta}}{2\zeta} \biggr) \,.  \label{eq:bump_z-zeta}
\end{eqnarray}
where $r_0$ is a constant, $a=r_0 \cosh \theta$ and $b= r_0 \sinh \theta$. 
Substituting Eq.~(\ref{eq:bump_zeta-w}) into Eq.~(\ref{eq:bump_z-zeta}), 
we obtain the mapping function between the $w$- and $z$-plane: 
\begin{eqnarray}
z = a \biggl( w + \frac{b}{a} \sqrt{w^2-1} \biggr) \equiv F(w) \,. \label{eq:bump_z-w}
\end{eqnarray}
The complex potential on the $w$-plane is given by 
\begin{eqnarray}
\widetilde{\Phi}(w) = -\widetilde{H}_0 w \,,  \label{eq:tildePhi}
\end{eqnarray}
which certainly yields the magnetic field on the $w$-plane, $H_u - iH_v = -\widetilde{\Phi}_{\rm bump}'(w) = \widetilde{H}_0$.   
Then the $\beta$-factor is derived from the general formula given by Eq.~(\ref{eq:beta}):  
\begin{eqnarray}
\beta(x,y) = \frac{1}{H_0} \biggl| \frac{\widetilde{H}_0}{a+bw/\sqrt{w^2-1}} \biggr| 
= \frac{1+\frac{b}{a}}{1+\frac{b}{a}\frac{w}{\sqrt{w^2-1}}} \,.  \label{eq:beta_bump}
\end{eqnarray}
where a condition $\beta \to 1\,(z\to \infty)$, namely, $\widetilde{H}_0 = (a+b) H_0$ is used. 
The maximum value of $\beta$ is given by the top of the bump ($z=ib$ or $w=0$):  
\begin{eqnarray}
\beta^* = \beta(0,b)=\beta|_{w=0} = 1+\frac{b}{a} \,. 
\end{eqnarray}
Note that $\beta^*|_{a=b}=2$ is consistent with the well-known results of $\beta$ of the cylinder.  

\section{2D pit with sharp edges review}
%
\begin{figure}[tb]
   \begin{center}
   \includegraphics[width=0.8\linewidth]{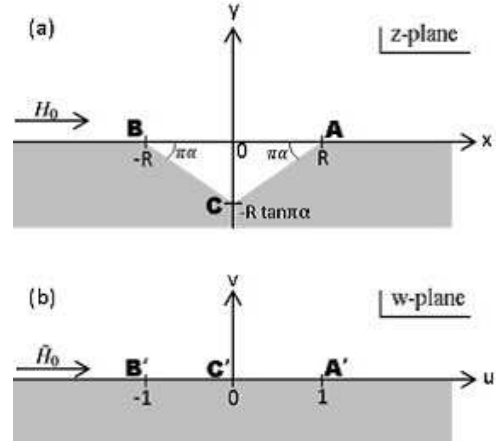}
   \end{center}\vspace{-0.2cm}
   \caption{
(a) 2D pit with sharp edges and its maps on (b) the $w$-plane. 
Gray regions correspond to superconductors in the Meissner state.  
A surface magnetic field far from the bump is given by $(H_x,H_y)=(H_0,\, 0)$ and $(H_u,H_v)=(\widetilde{H}_0,0)$ on the $z$- and $w$-plane, respectively. 
   }
\label{fig2}
\end{figure}

Let us review a model with sharp edges shown in Fig.~\ref{fig2}. 
The map $z=F(w)$ is given by the Schwarz-Christoffel transformation: 
\begin{eqnarray}
z= F(w) = \kappa R \int_0^w \!\! f(w')  dw' + \lambda R  \, ,  \label{eq:SC}  
\end{eqnarray}
where the integrand $f(w)$ is given by
\begin{eqnarray}
f (w) = \frac{(w^2-1)^{\alpha}}{w^{2\alpha}} \,.   \label{f0} 
\end{eqnarray}
The constant $\lambda$ is determined by the condition that $\rm C'$ on the $w$-plane is mapped into $\rm C$ on the $z$-plane: 
\begin{eqnarray}
\lambda = -i \tan(\pi \alpha) \,. 
\end{eqnarray}
Similarly the constant $\kappa$ is determined by the condition that $\rm A'$ on the $w$-plane is mapped into $\rm A$ on the $z$-plane: 
\begin{eqnarray}
\kappa = \frac{\sqrt{\pi}}{\cos{\pi\alpha} \, \Gamma(1+\alpha)\Gamma(\frac{1}{2}-\alpha)} \, . \label{eq:kappa_sharp}
\end{eqnarray}
The complex potential on the $w$-plane is given by 
\begin{eqnarray}
\widetilde{\Phi}(w) = -\widetilde{H}_0 w \,,  
\end{eqnarray}
Then $\beta$ is derived from the general formula given by Eq.~(\ref{eq:beta}):  
\begin{eqnarray}
\beta(x,y)=\frac{1}{H_0} \biggl|\frac{\widetilde{H}_0}{\kappa R f(w)}\biggr| = \frac{1}{|f(w)|}
= \biggl| \frac{w^{2\alpha}}{(w^2-1)^{\alpha}} \biggr| \,,
\end{eqnarray}
where a condition $\beta \to 1\,(z\to \infty)$, namely, $\widetilde{H}_0 = \kappa R H_0$ is used. 
The $\beta$ at $\rm A$, $\rm B$ and $\rm C$ are immediately obtained as
\begin{eqnarray}\label{eq:betaABC}
  \begin{cases}
    \beta({\rm A}) = \beta|_{w=1}  = \infty   \,, \\
    \beta({\rm B}) = \beta|_{w=-1}  = \infty   \,, \\
    \beta({\rm C}) = \beta|_{w=0}  = 0 \, .
  \end{cases}
\end{eqnarray}
The magnetic field is enhanced at the edges, 
vanishes at the bottom, 
and is not enhanced far from the pit. 
The expansion in the vicinity of the edges and its extrapolation to the model with round edges are found in Ref.~\cite{Kubo}. 

\section{2D pit with chamfered edges}

%
\begin{figure}[t]
   \begin{center}
   \includegraphics[width=0.8\linewidth]{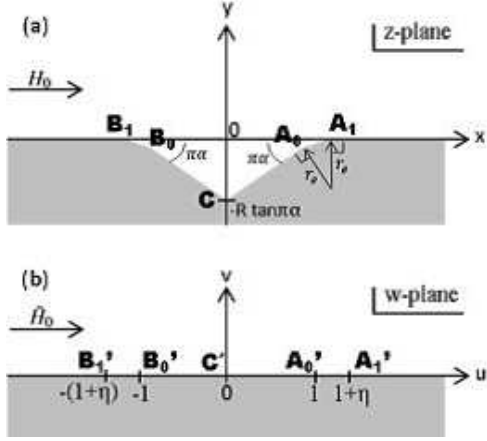}
   \end{center}\vspace{-0.2cm}
   \caption{
(a) 2D pit with chamfered edges and its maps on (b) the $w$-plane. 
Gray regions correspond to superconductors in the Meissner state.  
A surface magnetic field far from the bump is given by $(H_x,H_y)=(H_0,\, 0)$ and $(H_u,H_v)=(\widetilde{H}_0,0)$ on the $z$- and $w$-plane, respectively. 
   }
\label{fig3}
\end{figure}

A pit with chamfered edges shown in Fig.~\ref{fig3} can be regarded as an approximation of a pit with round edges. 
The map $z=F(w)$ is given by the Schwarz-Christoffel transformation with 
\begin{eqnarray}
f(w) = \frac{ \{ w^2 -(1+\eta)^2 \}^{\frac{\alpha}{2}}  (w^2-1)^{\frac{\alpha}{2}} }{w^{2 \alpha}} \,. 
\end{eqnarray}
where $\eta$ is a constant. 
The constant $\lambda$ is determined by much the same way as 2D pit with sharp edges. 
The condition that $\rm C'$ on the $w$-plane is mapped into $\rm C$ on the $z$-plane yields
\begin{eqnarray}
\lambda = -i \tan\pi \alpha \,. 
\end{eqnarray}
The constant $\kappa$ is determined by the condition that $\rm A_0'$ on the $w$-plane is mapped into $\rm A_0$ on the $z$-plane: 
\begin{eqnarray}
&& R +r_e\tan\frac{\pi\alpha}{2} -r_e\sin\pi\alpha -ir_e(1-\cos\pi\alpha) \nonumber \\
&& = F(1) 
= \kappa R \int_0^1 \!\! f(w)  dw - i R\tan\pi\alpha \, . \nonumber 
\end{eqnarray}
Using $1+i\tan\pi\alpha =e^{i\pi\alpha}/\cos\pi\alpha$, 
$1+i\tan(\pi\alpha/2)=e^{i\frac{\pi\alpha}{2}}/\cos(\pi\alpha/2)$ and 
$f(w) = e^{i\pi\alpha} ((1+\eta)^2-w^2)^{\frac{\alpha}{2}}(1-w^2)^{\frac{\alpha}{2}} w^{-2\alpha}$, 
the abvoe condition becomes
\begin{eqnarray}
&&\frac{1}{\cos \pi\alpha} -\frac{r_e}{R}\tan\frac{\pi\alpha}{2} \nonumber \\
&& =\kappa \int_0^1 \!\! dw \,  ((1+\eta)^2-w^2)^{\frac{\alpha}{2}}(1-w^2)^{\frac{\alpha}{2}} w^{-2\alpha}  \nonumber  \\
&& = \frac{\kappa}{2} \int_0^1 \!\! dt \, t^{-\alpha-\frac{1}{2}} (1-t)^{\frac{\alpha}{2}}  ((1+\eta)^2-t)^{\frac{\alpha}{2}}  \, ,  \nonumber \\ 
&& =\frac{\kappa}{2} (1+\eta)^{\alpha} \int_0^1 \!\! dt \, t^{a-1} (1-t)^{c-a-1}  (1-t\zeta)^{-b} \nonumber \\
&& = \frac{\kappa}{2} (1+\eta)^{\alpha} \frac{\Gamma(a)\Gamma(c-a)}{\Gamma(c)}\, {_2}F_1(a, b;  c, \zeta)\, ,
\nonumber  
\end{eqnarray}
where $t\equiv w^2$, $a\equiv -\alpha +\frac{1}{2}$, $b\equiv -\frac{\alpha}{2}$, $c\equiv -\frac{\alpha}{2} + \frac{3}{2}$, $\zeta \equiv (1+\eta)^{-2}$, and $_2 F_1 (a, b;  c, \zeta)$ is the Gaussian hypergeometric function. 
The gamma functions in the last line can be simplified by using Legendre's duplication formula, $\Gamma(1+\frac{\alpha}{2}) \Gamma(\frac{\alpha}{2}+\frac{1}{2}) =2^{-\alpha}\sqrt{\pi} \Gamma(\alpha+1)$, and Euler's reflection formula, $\Gamma(\frac{1}{2}-\frac{\alpha}{2})\Gamma(\frac{1}{2}+\frac{\alpha}{2})=\pi/\sin(\frac{\pi}{2}-\frac{\pi\alpha}{2}) = \pi/\cos\frac{\pi\alpha}{2}$. 
Then the condition is finally written as 
\begin{eqnarray}
&&\frac{1}{\cos \pi\alpha} -\frac{r_e}{R}\tan\frac{\pi\alpha}{2} \nonumber \\
&&= \kappa \frac{(1+\eta)^{\alpha} \cos\frac{\pi\alpha}{2} \Gamma(\frac{1}{2}-\alpha) \Gamma(1+\alpha)}{2^{\alpha} \sqrt{\pi} (1-\alpha)} \, {_2}F_1(a, b;  c, \zeta)\, , \nonumber 
\end{eqnarray}
namely, 
\begin{eqnarray}
\kappa \!=\! 
\frac{1-\frac{r_e}{R} \cos\pi\alpha \tan\frac{\pi\alpha}{2}}{(1+\eta)^{\alpha}} 
\frac{2^{\alpha}\,(1-\alpha)\, \kappa_0}{ {_2}F_1(a, b;  c, \zeta) \cos\frac{\pi\alpha}{2}}  , 
\label{eq:kappa_chamfered}
\end{eqnarray}
Note here that Eq.~(\ref{eq:kappa_chamfered}) converges Eq.~(\ref{eq:kappa_sharp}) as $r_e/R \to 0$.  
The constant $\eta$ is determined by the condition that $\rm A_1'$ on the $w$-plane is mapped into $\rm A_1$ on the $z$-plane: 
\begin{eqnarray}
R +r_e\tan\frac{\pi\alpha}{2} =F(1+\eta) = F(1) + \kappa R \int_1^{1+\eta} \!\!\!\!dw\, f(w) \,. \nonumber 
\end{eqnarray}
Substituting $f(w)=(-1)^{\frac{\alpha}{2}} \{ (1+\eta)^2 - w^2 \}^{\frac{\alpha}{2}} (w^2-1)^{\frac{\alpha}{2}} w^{-2\alpha}$, the above condition becomes
\begin{eqnarray}
2 \frac{r_e}{R} \sin\frac{\pi\alpha}{2} 
\!=\! \kappa \int_1^{1+\eta} \!\!\!\!\!\!\!\!dw\, \frac{ \{ (1+\eta)^2 -w^2\}^{\frac{\alpha}{2}} (w^2-1)^{\frac{\alpha}{2}} }{ w^{2\alpha} } .  \nonumber 
\end{eqnarray}
The integral can be performed under the assumption $\eta \ll 1$.  
Substituting $w=1+\epsilon\,(\epsilon \ll 1)$, we find
\begin{eqnarray}
2 \frac{r_e}{R} \sin\frac{\pi\alpha}{2} 
= \frac{2^{\alpha+1} \kappa}{2+\alpha} \eta^{\alpha+1}
, \nonumber 
\end{eqnarray}
or, 
\begin{eqnarray}
\eta = \biggl( \frac{(2+\alpha) \sin\frac{\pi\alpha}{2}}{2^{\alpha} \kappa} \frac{r_e}{R}\biggr)^{\frac{1}{1+\alpha}}
,  \label{eq:eta}
\end{eqnarray}
Then $\beta$ is derived from the general formula given by Eq.~(\ref{eq:beta}):  
\begin{eqnarray}
\beta(x,y)= \frac{1}{|f(w)|}
= \biggl|  \frac{w^{2 \alpha}}{ \{ w^2 -(1+\eta)^2 \}^{\frac{\alpha}{2}}  (w^2-1)^{\frac{\alpha}{2}} }   \biggr| \,, 
\label{eq:beta_pit_chamfered}
\end{eqnarray}
with Eq.~(\ref{eq:eta}) and (\ref{eq:kappa_chamfered}). 

\section{summary}
Formulae of $\beta$ at a 2D semi-elliptical bump and a 2D pit with chamfered edges are given by Eq.~(\ref{eq:beta_bump}) and (\ref{eq:beta_pit_chamfered}), respectively. 
See Ref.~\cite{Kubo} for the pit with round edges.

\end{document}